# Location-Based Output Adaptation for Enhanced Actuator Performance using Frequency Sweep Analysis

Kevin Fischler[1], Seungjae Oh[2], and Soekhee Jeon[3]

[1, 3] Department of Computer Engineering, Kyung Hee University, Yongin-si, South Korea

[2] Department of Software Convergence, Kyung Hee University, Yongin-si, South Korea

(Email: jeon@khu.ac.kr)

**Abstract ---** This paper presents a methodology for enhancing actuator performance in older devices or retrofitting devices with haptic feedback actuators. The approach is versatile, accommodating various actuator and mounting positions. Through a frequency sweep analysis, the system's characteristics are captured, enabling the creation of location-specific transfer functions to accurately transform input signals into command signals for a precise output at the target location. This method offers fast and simple collection of the system properties and generation of location-specific signals.

## 1 Introduction

Many smartphones and other touch devices are equipped with some form of haptic sensations (e.g., vibrotactile feedback). However, the poor performance or inaccurate feedback in many devices with existing vibrotactile feedback systems often stems from the inherent limitations of the actuators or actuator types used [1].

In particular, there are two main causes for this issue. First, the haptic feedback does not account for rendering based on interaction parameters (e.g., the applied speed). Second, the devices are not able to adapt to the natural frequency response of the attached actuators and their feedback at the target location, which can affect the quality and transparency of the haptic feedback. This paper focuses on the latter.

Thus, to improve the rendering quality and account for actuators' natural frequency response along with the target interaction location, we propose a simple and efficient method to deliver more accurate haptic feedback. We begin with system identification, followed by the estimation of the transfer function of the system at the target location. Using the estimated system properties, a command signal can be generated that takes these properties into account. This process of estimating the system properties is repeated until an accurate system estimation is achieved for every target location. It is noted that the overall approach is inspired by [2] and is briefly discussed in Sect. 2. We believe that the proposed approach can provide an additional level of information through haptic-based sensations, such as simple notifications [3] or even rendering of high-frequency textures [4] accurately on the desired location. This approach would enable accurate signal reproduction even on larger devices and in cases where the actuator position and touch location are farther apart. Below we detail the proposed method followed by the hardware setup used.

## 2 Methodology

### 2.1 Workflow

The dynamic properties of the system are analyzed by conducting a frequency sweep. The resulting output signal is used to derive a transfer function for system modeling [2], [5]. In contrast to the methodology outlined in [2], this paper considers the system properties of both the actuator and the touch device, as well as the position of the measurement. For accurate reproduction based on touch location, transfer functions are estimated for each target location. The output signal from the actuator is captured using an acceleration sensor. To streamline and expedite this procedure, a MATLAB application has been developed. The application guides the recording process, processes the signal, and creates the transfer function at each target location. The workflow of the application is shown in Fig. 1. System Identification at Target

Location: A frequency sweep is performed. Estimation of Transfer Function: A system response is recorded, and the transfer function is estimated. Adaptation of Input Signal to Command Signal: The transfer function is applied to the input signal to create a command signal. Playback of Command Signal: The command signal is sent to the actuator. Updating of Transfer Function at Target Location: Conducting a new frequency sweep and/or higher-order transfer function estimation. Change Target Location: Placing the acceleration sensor at a new location. The working flow concept is based on [2].

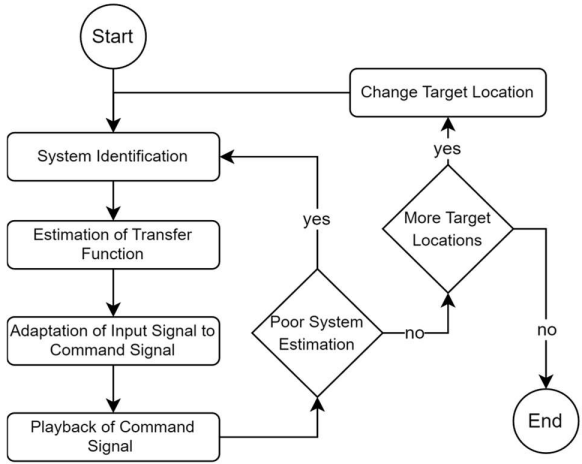

Fig. 1. Workflow of recording and collecting system properties per target location.

### 2.2 Hardware setup

The hardware setup consists of a Microsoft Surface PC, a TPA3116D2 audio amplifier, a NI DAQ, an ADXL335 3-axis accelerometer and a TacHammer Carlton haptic actuator. The brief overview of the system architecture can be seen in Fig. 2 while the Fig 3. displays the actual sensor and actuator position on the touch device.

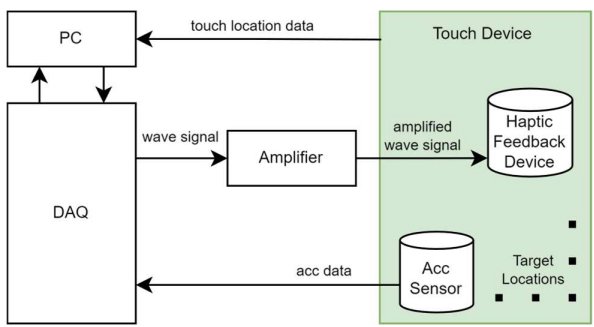

Fig. 2. Hardware setup for recording and estimation of the system properties of the touch device. The target location is set by the acceleration sensor position.

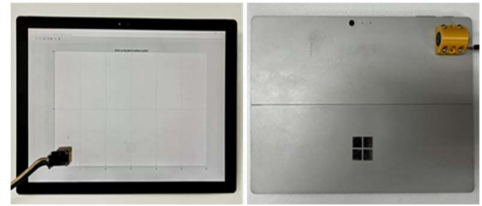

Fig. 3. Left: Front of the touch device with the acceleration sensor located on the bottom left-hand side of the screen. Right: Backside of the touch device with the actuator positioned on the upper right-hand side.

## 3 CONCLUSION

The concept presented aims to deliver a fast and easy method to retrofit touch devices with haptic feedback. While simplicity is a defining characteristic of this approach, it is imperative to recognize that any changes in the device's environment (such as holding position or mounting position) post-recording necessitate the estimation of new systems properties. The methodology could be expanded by incorporating interpolation functions to more accurately describe the space between measured locations. Additionally, there is potential for the implementation of alternative approaches to describe the system properties.

### ACKNOWLEDGEMENT

This research was supported by the IITP under the Ministry of Science and ICT Korea through the IITP program No. 2022-0-01005 and under the metaverse support program to nurture the best talents (IITP-2024-RS-2024-00425383).

### REFERENCES

[1] Choi, S. and Kuchenbecker, K.J., 2012. Vibrotactile display: Perception, technology, and applications. Proceedings of the IEEE, 101(9), pp.2093-2104.

[2] Ujitoko, Y., Sakurai, S. and Hirota, K., 2020, March. Vibrator transparency: Re-using vibrotactile signal assets for different black box vibrators without re-designing. In 2020 IEEE Haptics Symposium (Haptics) (pp. 882-889). IEEE.

[3] Qian, H., Kuber, R. and Sears, A., 2013. Tactile notifications for ambulatory users. In CHI'13 Extended Abstracts on Human Factors in Computing Systems (pp. 1569-1574).

[4] Kyung, K.U. and Lee, J.Y., 2008. Ubi-Pen: A haptic interface with texture and vibrotactile display. IEEE Computer Graphics and Applications, 29(1), pp.56-64.

[5] McMahan, W. and Kuchenbecker, K.J., 2014, February. Dynamic modeling and control of voice-coil actuators for high-fidelity display of haptic vibrations. In 2014 IEEE Haptics Symposium (HAPTICS) (pp. 115-122). IEEE.